\documentclass[
 reprint,
 aps,
 superscriptaddress
]{revtex4-1}

\usepackage{graphicx}

\begin{document}

\title{Electric moulding of dispersed lipid nanotubes\\into a nanofluidic device
}
\author{Hiroshi Frusawa}
\email{frusawa.hiroshi@kochi-tech.ac.jp}
\author{Tatsuhiko Manabe}
\author{Eri Kagiyama}
\affiliation{Institute for Nanotechnology, Kochi University of Technology, Tosa-Yamada, Kochi 782-8502, Japan.}
\author{Ken Hirano}
\affiliation{Health Research Institute, National Institute of Advanced Industrial Science and Technology (AIST), 2217-14, Hayashi-cho, Takamatsu, Kagawa, 761-0395, Japan.}
\author{Naohiro Kameta}
\author{Mitsutoshi Masuda}
\affiliation{Nanosystem Research Institute, National Institute of Advanced Industrial Science and Technology (AIST), Tsukuba Central 5, 1-1-1 Higashi, Tsukuba, Ibaraki 305-8565, Japan.}
\author{Toshimi Shimizu}
\affiliation{AIST Fellow, National Institute of Advanced Industrial Science and Technology (AIST), Tsukuba Central 5, 1-1-1 Higashi, Tsukuba, Ibaraki 305-8565, Japan.}


\begin{abstract}
Hydrophilic nanotubes formed by lipid molecules have potential applications as platforms for chemical or biological events occurring in an attolitre volume inside a hollow cylinder.
Here, we have integrated the lipid nanotubes (LNTs) by applying an AC electric field via plug-in electrode needles placed above a substrate.
The off-chip assembly method has the on-demand adjustability of an electrode configuration, enabling the dispersed LNT to be electrically moulded into a separate film of parallel LNT arrays in one-step.
The fluorescence resonance energy transfer technique as well as the digital microscopy visualised the overall filling of gold nanoparticles up to the inner capacity of an LNT film by capillary action, thereby showing the potential of this flexible film for use as a high-throughput nanofluidic device where not only is the endo-signalling and product in each LNT multiplied but also the encapsulated objects are efficiently transported and reacted.
\end{abstract}

\maketitle

A large research effort has been made to organise colloidal particles into functional materials, and we now have a variety of elaborate techniques for engineering colloids including functional nanoparticles \cite{xia,colloid-rev,direct-rev,velev1,yeth,van}.
Colloidal methods guide particle configurations in desired directions using the bottom-up phenomena of either self-organisation \cite{xia,colloid-rev,direct-rev,velev1,yeth} or directed assembly using external fields \cite{direct-rev,velev1,yeth,van}.
Polymerized crystalline arrays of spherical colloids represent the bottom-up assembly into macroscopic or mesoscopic materials, which has been found to be available for use in photonic devices \cite{xia,colloid-rev,nature-gel}.
Furthermore, the assembly materials serve for the efficient utilisation of the constituents' individual properties.

Among the functional nanoparticles of various shapes, we focus on lipid nanotubes (LNTs) whose tubular structures include those with hydrophilic internal and external membrane surfaces formed by lipid molecules \cite{schnur,zhang, orwar1,orwar2,kamiya,shimizu1,shimizu2,shimizu3,masuda}.
The LNT used has a typical 10-nm inner diameter and 10-$\mu$m length \cite{kamiya,shimizu1,shimizu2,shimizu3,masuda}, and therefore provides the confined liquid space of ten attolitre volume, which is smaller than the volumes of a femtolitre chamber and a single cell by factors of $10^2$ and $10^5$, respectively \cite{shimizu1,shimizu2}.
The confined liquid nanospace has found potential in chemical and biological applications of LNTs as well as polymer nanotubes \cite{masuda,martin,wakasugi,ding,karam,losic1,chaperone}, including controlled drug release \cite{wakasugi,ding,karam,losic1} and artificial chaperoning of denaturated proteins \cite{chaperone}.
An individual LNT has thus been found to be useful as a nanoreactor and/or nano-assay device \cite{kameta2007,orwar3,nat-bio,label-free,droplet,manz-review}.
Correspondingly, a variety of techniques has been developed for manipulating and integrating LNTs into ordered nanocapillary arrays \cite{frusawa1,guo,fujima,arai,hirano,fang1,fang2,fang3}.

Despite the numerous attempts, there remain unresolved issues in creating nanofluidic devices of LNT arrays \cite{losic1,fang1,fang2,fang3,japanese,li-koshizaki,losic2,tan,yaman}.
Here we address the methodological challenges associated with the following: (i) sufficient total output, (ii) efficient transport and reaction inside the hollow cylinder, and (iii) no addition of binder molecules to keep the native outer surfaces.

%%%%%%%%% FIG1 %%%%%%%%%%%%%%
\begin{figure*}[hbtp]
        \begin{center}
	\includegraphics[
	width=13.5cm
	]{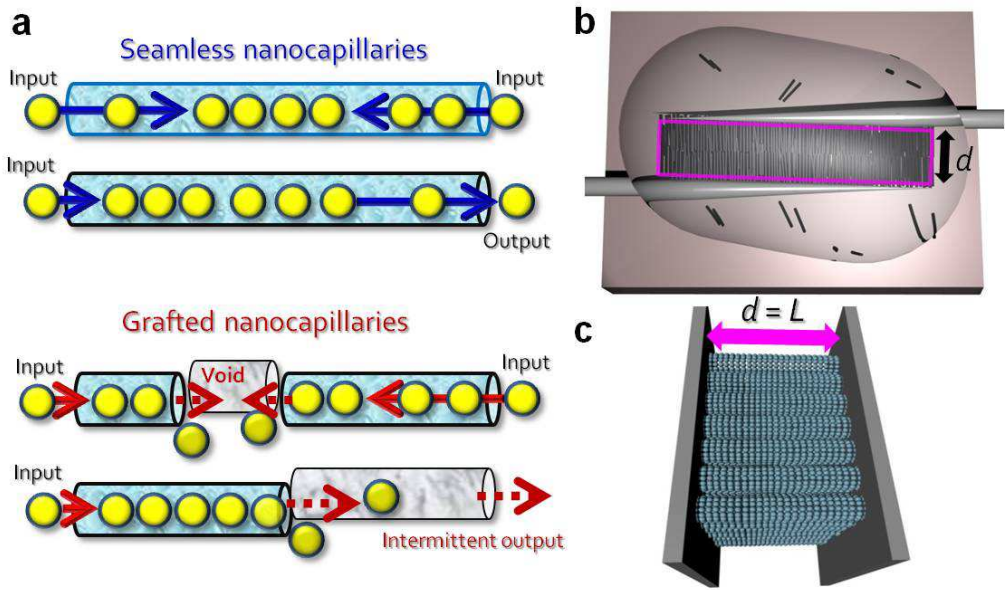}
	\end{center}
	\caption{{\bf Off-chip assembly of dispersed LNTs into parallel nanocapillary arrays}.
({\bf a}) Comparison between seamless and grafted nanocapillaries.
The upper situation illustrated by two schematics represents seamless nanocapillaries that have high-throughput processing.
The lower diagrams correspond to instances of inefficient transportations and reactions, which demonstrates that the graft-induced jamming of injected particles either stops the flows out of the nanotubes or causes incomplete immersion of the inner space, leaving void nanotubes.
({\bf b}) Illustration of the plug-in system.
An external AC electric field is locally applied to a drop of an LNT suspension by inserting a pair of electrode needles controlled by patch-clamp micromanipulators into the solution.
The purple rectangle marks the region for fundamental studies of the assembly processes.
({\bf c}) Schematic representation of parallel nanoapillary arrays that are formed by sandwiched LNTs when the electrode gap $d$ is adjusted to the mean long axial length $L$ of a single LNT.
}
\end{figure*}
%%%%%%%%%%%%%%%%%%%%%

(i) First, the increase in the total output requires the cost-effective assembly of a considerable number of nanocapillaries into the oriented arrays.
We need $10^5$ LNTs for the output to be equivalent to that of a single cell.
Therefore, it is natural to highlight the bottom-up assembly techniques instead of the top-down approaches.
(ii) Second, a high-throughput device requires that open ends are connected by a seamless nanocapillary inside a single LNT (see Fig. 1(a)), which is similar to the ideal state of semiconducting carbon nanotubes in creating a thin film transistor \cite{carbon1,carbon2}. 
This bridging can be accomplished by excluding serial arrays of LNTs from the oriented assembly.
(iii) Last, a genuine assembly with an unaltered native outer surface is vital to the straightforward scale-up of a pilot study on a single LNT, a nano-platform for reacting and transporting encapsulated objects.

\subsection*{Results}
{\bf Off-chip assembly for creating nanofluidic devices}. 
In order to fabricate a flexible nanofluidic device of LNTs fulfilling the aforementioned requirements, we exploited an AC electric field to direct the organisation of nanoparticles \cite{velev1,yeth,van,carbon1,carbon2,velev2,velev3,chaikin1,chaikin2,juarez,ellipsoid,voldman}.
While previous studies on the electric assembly of colloidal spheres have demonstrated that AC fields applied via on-chip electrodes precisely control the 3D and 2D crystalline orientations as well as the colloidal concentrations \cite{velev1,yeth,van,velev2,velev3,chaikin1,chaikin2,juarez}, on-chip assembly systems have problems arising from the electrodes being fixed on a substrate.
Because the electrode configuration cannot be changed, on-chip systems are irrelevant  to electrode removal from a colloidal assembly without the use of a binder.
Additionally, on-chip processes depend on the wettability of the substrates and limit the substrates that can be used for applications.

To overcome these problems, we adopt the off-chip assembly system that plugs into a sample drop a pair of electrode needles above a cover slip (see Fig. 1(b)).
The rectangle delineated in Figure 1(b) corresponds to an off-chip potential valley created by the needle pair.
Because the plug-in electrodes are independently controlled by two sets of patch clamp micromanipulators, the off-chip system facilitates to adjust the electrode gap so that serial nanocapillary alignments may be excluded, thereby creating an LNT film of parallel nanocapillary arrays (see Fig. 1(c)).
Moreover, the off-chip style has advantages of avoiding both impurity deposition and drying-induced adhesion to either the electrodes or the substrate.
Therefore, we are able not only to control the assembly size, but also to remove an assembly from the needles, which yields a separated genuine assembly of an appropriate size.
We can see the assembly and drying sequences for the LNT suspension in Supplementary Movies 1 and 2, respectively.

%%%%%%%%% FIG2 %%%%%%%%%%%%%%
\begin{figure*}[htbp]
        \begin{center}
	\includegraphics[
	width=13cm
	]{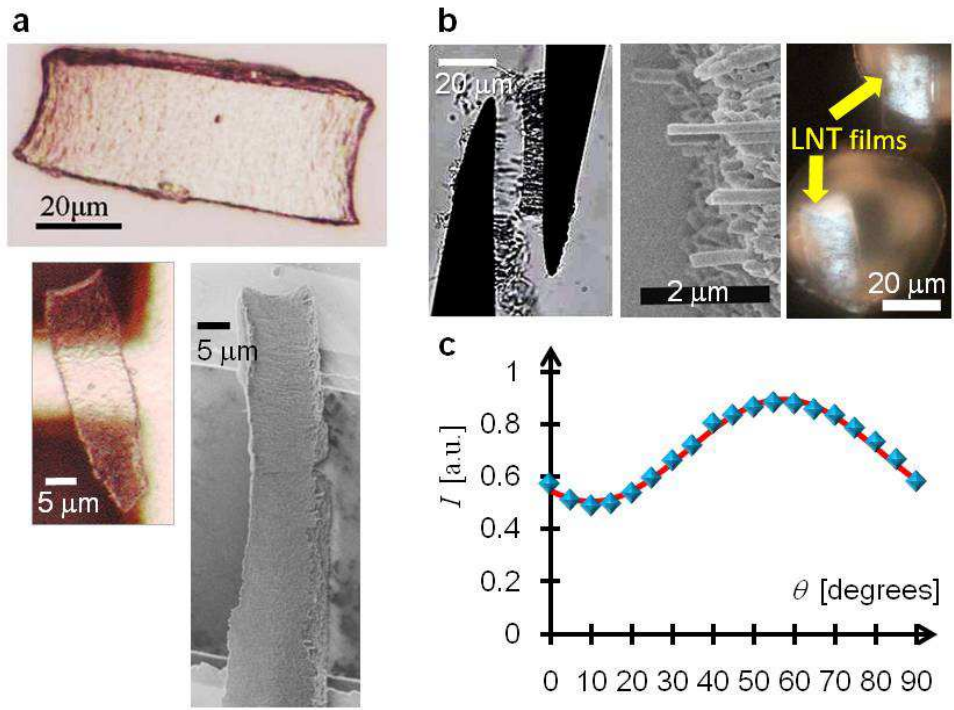}
	\end{center}
	\caption{{\bf Anisotropy of electrically moulded LNT films}.
({\bf a})
A digital microscope image on the upside shows a film of LNT arrays successfully removed from the electric mould with a width and length of 20 $\mu$m and 70 $\mu$m, respectively.
The lower images, on the other hand, represent narrower LNT films having a width of approximately 10 $\mu$m.
These LNT films are placed on a metallic grid substrate, on which in situ observations are performed using digital microscopy (left image) and SEM (right image).
({\bf b}) 
The leftmost micrograph shows an instant the needles were removed from the dried assembly.
The middle SEM image provides a magnified view of an edge, which verifies that LNTs are aligned in the desired direction.
The birefringence due to the LNT alignment is seen from the right image, the polarised optical micrograph using cross Nicols, where we can find a pair of bright LNT films on the SEM grid.
({\bf c})
The blue squares show the angular dependence of the mean brightness evaluated using polarised microscopy when the sample stage is successively rotated. The red sine curve is fitted to the data.
}
\end{figure*}
%%%%%%%%%%%%%%%%%%%%%

The LNT length distribution has been efficiently determined using in situ measurements of the rotational dynamics in the LNT suspension (see Supplementary Discussion for the details) \cite{hirano}.
The histogram in Supplementary Figure S1 shows that the majority of the lengths are between four and six microns;
the mean length is calculated to be approximately $6\,\mu\mathrm{m}$.
It is also to be noted that the off-chip assembly was performed in a filtered dispersion where shorter LNTs had passed away through a 5-$\mu\mathrm{m}$ plastic filter (see Methods).
Accordingly, an electrode gap less than $10\,\mu\mathrm{m}$ can considerably diminish the number of serial-parallel arrays, the alignment of two or more LNTs along the external electric field, so that we can obtain an LNT film whose majority is formed by parallel nanocapillaries.
In what follows, we demonstrate that this off-chip technique using plug-in electrodes allows one-step fabrication of biocompatible nanofluidic devices solely using LNTs, which will be referred to as "{\itshape electric moulding}".

{\bf Electric moulding into birefringent LNT films}. 
Figure 2(a) exhibits three typical images of LNT films.
A digital microscope image on the upside shows a spontaneously dried film on a cover glass, a substrate used in separating LNT assembly from manipulated electrodes (the last procedure of electric moulding).
While the LNT film in the upper image, which leaves moist, has a 20-$\mu$m width and 70-$\mu$m length in accordance with the electrode configuration, it is possible to change the size of the electrically focused area into 10-$\mu\mathrm{m}$ in gap length which satisfies the aforementioned requirement of parallel LNT arrays.
The lower images in Figure 2(a) represent the narrower LNT films having 10-$\mu\mathrm{m}$ width. 
These films are placed on a grid substrate to perform in situ observations using scanning electron microscopy (SEM) as shown in the lower right image, and the metallic substrate can be seen through a transparent LNT film in the digital microscope image on the lower left side.

%%%%%%%%% Figure3 %%%%%%%%%%%%%%
\begin{figure*}[hbtp]
        \begin{center}
        \includegraphics[width=13cm]{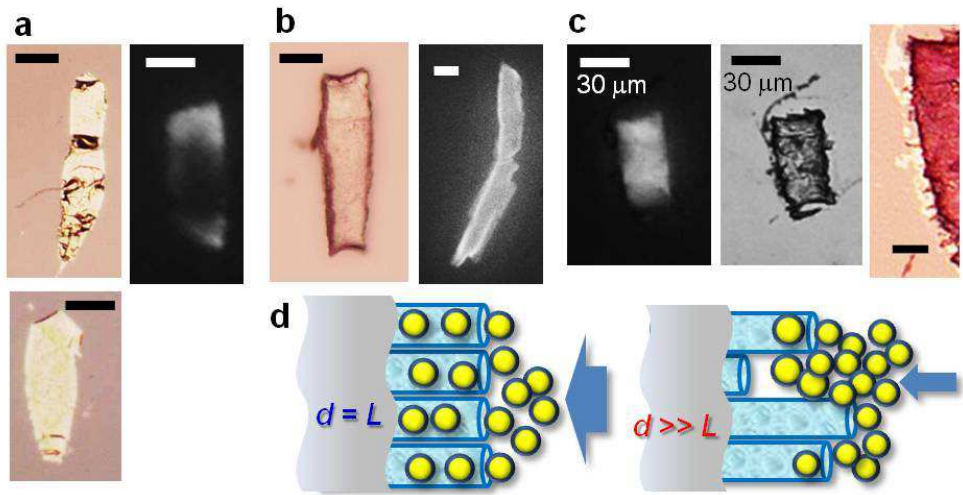}
	\end{center}
	\caption{{\bf Addition of gold nanoparticles to LNT films}.
All of the scale bars unspecified represent 10 $\mu$m.
({\bf a})
The upper and lower images on the left side are digital micrographs of typical GNP-included films having a width of approximately 10 $\mu$m, which have been mixed with 5-nm GNPs.
From a fluorescence micrograph on the right side, we can see a partially fluorescent part on a border area of a GNP-included film stained by R6G fluorescent molecules.
({\bf b})
While a digital microscopy image on the left side shows an LNT film mixed with 40-nm GNPs, a fluorescence micrograph on the right side exhibits a bright rectangle, which corresponds to an R6G-added film that has been mixed with 40-nm GNPs.
({\bf c})
Despite the mixture of 5-nm GNPs, an R6G-added film having 30-$\mu$m width provides a fluorescent area overall the film (the leftmost image), whereas the middle image observed using transmitted light displays the identical film and the digital microscopy magnifies the edge (the rightmost micrograph).
({\bf d})
A schematic comparison between even and rough ends in capillary action of GNPs into LNT films.
The sectional plane is more likely to be flat as the width $d$ is closer to the mean LNT length $L$, facilitating to capture GNPs as shown in the left schematic.
The right diagram, on the other hand, illustrates that an irregular entrance formed by uneven terminations on the edge interferes with GNP-encapsulation into the inner space of individual LNTs.
}
\end{figure*}
%%%%%%%%%%%%%%%%%%%%%%%%%%%%%%%

Electrode removal does not always succeed, and the left micrograph in Figure 2(b) demonstrates a typical tearing situation during the separation process from the electrodes;
one of the methodological refinements for reducing incomplete removal would be to improve the electrode surface using chemical modification.
At the same time, the left image suggests that each individual LNT is perpendicular to the surface of electrode needles, which has been verified by the middle SEM image in Figure 2(b).
The SEM image focuses on an edge parallel to the electrode surface, showing the alignment of LNTs.
The polarised optical micrograph on the rightmost side of Figure 2(b) also demonstrates the birefringent property of LNT films placed on the SEM grid.
The film brightness measured in the polarised micrograph varies when the stage is rotated.
Image analysis provides the intensity of the transmitted light, normalised by the maximum, which is then averaged over the regions of these films to obtain a mean brightness $I(\theta)$ as a function of the rotation angle $\theta$ of the stage (see Fig. 2(c)).
The fitting result is given by $I(\theta)=A_0+A\sin\left\{2\pi(\theta/\Theta)+\delta\right\}$ with a set of optimized parameters, $A_0=0.7$, $A=0.19$, $\Theta=90^\circ$, and $\delta=-129^\circ$, quantifying the anisotropic optical property common to liquid crystalline materials.

{\bf Size-selective inclusion of GNPs into LNT films}. The birefringent film of aligned LNTs is expected to selectively and efficiently capture aqueous nanoparticles by capillary action, when the target film is dried in a vacuum.
We demonstrate below that the LNT films fabricated by our electric moulding technique hold this capability of individual LNTs, using both the digital microscopy and the fluorescence resonance energy transfer (FRET) technique.
It has been found that the FRET solely occurs from the internal fluorescent molecules of Rhodamine 6G (R6G) to GNPs encapsulated in the inner space of LNTs because the thickness of the LNT membrane wall is approximately 80 nm and is large enough to prevent FRET through the wall \cite{kameta2007};
therefore, the FRET is relevant to the detection of nanoparticle encapsulation inside the LNTs.

The use of 5-nm GNPs provides the digital microscope images of Figure 3(a) where we observed that the LNT films were kept metallic after rinsing GNPs repeatedly, which is similar to the micrographic look of metallic grid in Figure 2(a).
Correspondingly, the addition of R6G molecules mostly exhibits no fluorescence due to the FRET, or the disappearance of the fluorescence band, and a fluorescence micrograph on the right side of Figure 3(a) selects an incomplete film that is somehow recognizable due to a partially fluorescent part where the lack of GNPs results in the irrelevance of the FRET.
As seen from the digital microscopy image of Figure 3(b), on the other hand, an LNT film mixed with 40-nm GNPs exhibits a normal surface indistinguishable from that of a genuine LNT film, which conforms with a fluorescent brightness highlighting the outline in the fluorescence micrograph on the right side of Figure 3(b).
The entire absence of FRET in adding 40-nm GNPs offers some explanations as follows:
while most of 40-nm GNPs are excluded from the hollow channels of LNTs having its inner diameter of $40\pm 10$ nm, the saline rinses serve to eliminate GNPs coated through vacuum drying and the FRET on the surface is negligible even if the covered GNPs are left to some extent.

The contrasted results of Figures 3(a) and 3(b) are interpreted by the size-selectivity of LNT films;
however, it remains to be proved that the FRET in LNT films actually requires the 10-$\mu$m width, though the relation, $d<2L$, should reduce serial arrays of LNTs in comparison with a wider film such that $d>2L$.
For making the ambiguity clear, we controlled another type of LNT film that was mixed with the identical 5-nm GNPs but was extended to 30 $\mu$m in width (see the middle micrograph in Fig. 3(c));
the 5-nm GNPs could be inserted into LNT films irrespectively of the width, if the FRET mainly occurs due to unwrapped GNPs such as those existing in the interstices among LNTs.
As shown in the left micrograph of Figure 3(c), the threefold width provides fluorescent films to the contrary.
Furthermore, it is found from comparing fluorescent micrographs in Figures 3(a) to 3(c) that the fluorescent area in this 30-$\mu$m film of Figure 3(c) does not cover a whole rectangle with its edge being obscure.
In other words, the FRET seems to be relevant on edges of a 30-$\mu$m film, which is consistent with a magnified view of this film in Figure 3(c) on the rightmost side where a metallic deposition of 5-nm GNPs is observed on the bordered region.

These results of Figure 3(c) imply the other significant factor of adjusting the width, which is schematically illustrated by Figure 3(d):
the inner capacity of parallel LNTs is efficiently and homogeneously filled with 5-nm GNPs when the electrode gap length $d$, adjusted close to the mean LNT length $L$, creates a flat section as shown in the left diagram in Figure 3(d).
For irregular ends, on the other hand, the inflow of 5-nm GNPs seems to be blocked out on the entrance region of open ends (see the right schematic in Fig. 3(d)), which explains the jamming of GNPs on film edges as shown in Figure 3(c).

%%%%%%%%%%%%%%%%%%%%%%
{\bf A comparative study of the assembly courses}.
We need to validate that the off-chip assembly is mainly driven by dielectrophoresis for systematically searching an optimum set of electric parameters, such as the strength and frequency of the applied electric fields.
To this end, we first investigate the assembly processes of carboxylate-modified polystyrene (CM-PS) spheres with their diameter of 3 $\mu$m, instead of LNTs.
As shown in Figure 4(a), the off-chip assembly of 3-$\mu$m spheres produces a hexagonal packing with two-dimensional crystalline order, as do the on-chip techniques.

%%%%%%%%% FIG4 %%%%%%%%%%%%%%
\begin{figure}[hbtp]
\begin{center}
	\includegraphics[
	width=8cm
	]{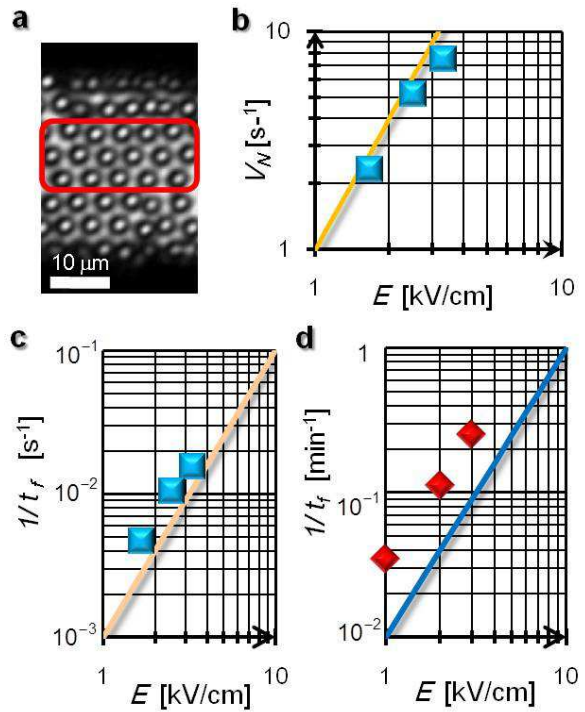}
	\end{center}
	\caption{{\bf Comparison between assembly rates of CM-PS particles and LNTs}.
({\bf a}) A magnified view of the first layer formed by CM-PS particles at $t=10$ seconds.
We can see hexagonal packing within the marked area.
({\bf b}) The assembly number rates $V_{N}$ of the spherical colloids as a function of $E$.
A straight line representing the relation $V_N\sim E^2$ is shown for comparison.
({\bf c},{\bf d}) Log-log plots of the inverse filling times $1/t_f$ versus $E$ regarding either CM-PS particles ({\bf c}) or LNTs ({\bf d}).
Straight lines in both ({\bf c}) and ({\bf d}), representing the relation $1/t_f\sim E^2$, are shown for comparison.
}
\end{figure}
%%%%%%%%%%%%%%%%%%%%%

As soon as the external field is applied at $t=0$ via the electrode pair, the 3-$\mu$m particles are attracted into the rectangular area, as depicted in Figure 1(b), and simultaneously we started to count the total number $N(t)$ of collected particles existing at a duration time $t$ in the rectangular region formed by the electrode pair, which has a 30-$\mu$m gap and a 90-$\mu$m width.
Incidentally, we have set the maximum total number in Supplementary Figure S2(a) to be less than $N_{\mathrm{max}}=7.4\times10^2$ because more electrically attracted colloids are localised around the electrode tips outside the rectangular region.
As evaluated in Supplementary Discussion, $N_{\mathrm{max}}$ is equal to that for randomly close packed spheres \cite{rcp} in the box of an electric holder having a minimum volume $A_{\mathrm{max}}H_{\mathrm{min}}$, with a prescribed rectangular area $A_{\mathrm{max}}=90\,\mu\mathrm{m}\times 30\,\mu\mathrm{m}=2700\,\mu\mathrm{m}^2$ and a lower bound of mean height $H_{\mathrm{min}}=6.1\,\mu\mathrm{m}$.
The former area is calculated from the aforementioned electrode configuration, whereas the latter height $H_{\mathrm{min}}$ reasonably explains the mean thicknesses of LNT films of several microns, which have been determined using the adjustment of the SEM focus when observing dry LNT films.

As shown in Supplementary Figure S2(a), the collected number, $N(t)$, as a function of the duration time $t$ is proportional to $t$ irrespective of the electric field strength in the range $0\leq N(t)\leq N_{\mathrm{max}}$, and therefore the global slope of $N(t)$ uniquely determines the assembly number rate, $V_N=dN(t)/dt$, per second.
Figure 4(b) is a log-log plot of $V_N$ versus the field strength $E$ at $f=100\,\mathrm{kHz}$, indicating that $V_N\sim E^{2}$ similarly to that of the dielectrophoretic velocity \cite{chaikin1,chaikin2,voldman}.
As explained in Supplementary Discussion, the $E$-dependence of $V_N$ proves the dielectrophoretic mechanism of the colloidal assembly in the plug-in system.

However, turning our attention to the LNTs, we find it hard to count individual LNTs, so an alternative parameter to $N(t)$ needs to be explored.
We then adopted the percentage fraction $P_A(t)$ of the cross-sectional area occupied by either CM-PS particles or electrically collected LNTs, where $P_A(t)$ is defined, using the relation $P_A(t)=100(A(t)/A_{\mathrm{max}})$, as the ratio of the measured area $A(t)$ to the maximum of $A_{\mathrm{max}}$.
Tracing the evolution of an occupied area that covers both the inside and outside of the rectangular region, $P_A(t)$ increases to and crosses 100 $\%$ at $t_f$ when colloidal particles with their number of $N(t_f)$ complete to fill the rectangular area prior to reaching the maximum state of $N_{\mathrm{max}}$ (see Supplementary Fig. S2(b)), and we have found that $N(t_f)$ of CM-PS particles is independent of $E$ as described in Supplementary Discussion.
Because $V_N$ is given by $V_N=N(t_f)/t_f$ using an invariant number of $N(t_f)$, it follows that the inverse of $t_f$ obeys the $E$-dependence of $V_N$.
Figure 4(c) actually demonstrates that $1/t_f\propto E^2$ for CM-PS particles in accord with the relation of $V_N$.

We then measured $P_A(t)$ using the optical micrographs to evaluate $t_f$ of LNTs (Supplementary Fig. S5(a)).
Figure 4(d) displays the log-log plot of $1/t_f$ versus $E$, where the experimental results satisfy the relation $1/t_f\sim E^2$ as well as that for CM-PS particles.
From Supplementary Figure S6, it is also found that the assembly degrees of CM-PS particles and LNTs are similarly reduced with an increase in $f$ in the range $100\,\mathrm{kHz}\leq f\leq 700\,\mathrm{kHZ}$, which corresponds to a typical $f$-dependence of dielectrophoresis \cite{hughes}.
Combining these consistent dependencies of the assembly courses of CM-PS particles and LNTs on $E$ and $f$, we conclude that dielectrophoresis causes the off-chip assembly of LNTs and of colloidal spheres.
Consequently, we fabricated the dry LNT films shown in Figures 2 and 3 using the electric parameters $E=3$ kV/cm and $f=100$ kHz, which have been optimised in terms of both the efficiency and robustness of the dielectrophoretic assembly.

The last parameter to be adjusted is the height $h$ from the substrate of the off-chip assembly in adopting an optimum set of $E=3$ kV/cm and $f=100$ kHz.
As described in Supplementary Discussion using the results of Supplementary Figure S4, we chose $h=50\,\mu\mathrm{m}$ for the fabrication of LNTs as a trade-off, considering the successful drying versus an efficient assembly or a sound removal of the electric mould.

\subsection*{Discussion}
To summarise, the off-chip assembly method using plug-in electrodes has three properties:
size adjustability of the potential valley, colloidal arrays lifted above the substrate, and electrode needles with tapered shapes.
The first two advantages of this technique have been described in detail, and we add another benefit associated with the last feature below.
The electrode tips produce the largest field gradient around them, thereby inducing colloidal dielectrophoresis mainly directed toward their pointed heads.
When we place the needle tips of twin absorbers on opposite sides as shown in Figure 1(b), colloidal flows are generated from the outside to the inside across both sides of the rectangular edges vertical to the electrodes (see Supplementary Fig. S3).
From the symmetrical pathways, it follows that gathering particles will meet in the middle region of the potential valley.
Similar flows during plug-in assembly of LNTs work to exclude voided assembly films caused by bottlenecks during the initial stage.

Final remarks are made on the potentials as nano-assay or nanoreactor devices.
Thanks to the reconfigurability, the off-chip assembly method is capable of providing a high-throughput nanofluidic device of LNTs that efficiently transports and reacts confined particles inside an LNT and that multiplies either the signalling or product of each LNT.
Moreover, the scalability of an LNT film to a macroscopic length opens up the possibility for creating a variety of flexible nanofluidic devices that make chemical or biological reactions visible.

\subsection*{Methods}

{\bf Materials}.
The lipid nanotubes (LNTs) used were derived from self-assembly in a glycolipid solution following a typical procedure, as described previously \cite{kamiya}.
The inner diameters were $40\pm 10$ nm, while the wall thicknesses were approximately 80 nm.
Because we set the electrode gap to 10 $\mu$m, the number of LNTs shorter than 5 $\mu$m, which are unfavourable particles, as described in the text (see also Figure 1(c)), needs to be diminished.
The following recipe was therefore adopted to prepare the LNT solution.
We first dissolved LNTs in deionised water to obtain a 0.1-wt.$\%$ dispersion, which was subsequently filtered through a 5-$\mu\mathrm{m}$ plastic filter (17594K, Sartorius).
Deionised water, 3-ml, was carefully added to the filter to redissolve the LNTs remaining on the filter.
A 20-$\mu$l aliquot of the filtered suspension was dropped onto a cover slip for electric assembly, yielding naturally dried LNT films.

The prepared LNT assembly was further dried in a vacuum to enhance the capillary action to fill the inner space of the LNTs with 5-nm GNPs.
We induced capillary action by dropping a 20-$\mu$l aliquot of a 0.1-wt.$\%$ GNP solution with 5-nm diameters (Wako) onto a dried LNT film.
Once again, we performed vacuum drying on the GNP-added films, which had subsequently been rinsed with an excess amount of 20-wt.$\%$ saline to remove GNPs from the outer surfaces of the LNTs.
To observe the FRET, we added a 0.1-mM solution of Rhodamine 6G (R6G, Wako) molecules to the film with encapsulated 5-nm GNPs.
As a control experiment, we also used a 0.1-wt.$\%$ solution of GNPs having a larger diameter of 40 nm (Wako).

For a reference study of the electric assembly process, we investigated the gathering courses of carboxylate-modified polystyrene (CM-PS) particles 3-$\mu$m in diameter (Dainichi-Seika) in a 100-$\mu$l aliquot of a 0.1 wt.$\%$ suspension.

\vspace{12pt}
{\bf Experimental setup}.
We mounted an aliquot on an inverted fluorescence microscope (Nikon, TE2000-U), into which a pair of tungsten microelectrodes with tip diameters of 1$\mu$m was inserted.
An external electric field having a sinusoidal wave was applied between the electrode pair using an arbitrary waveform generator (Agilent, 33220A) and a current amplifier (FLC Electronics, F30PV), and the microscopy images and movies were obtained using a CCD camera (Q-Imaging, Retiga Exi).
We performed video analysis using image analysis software (ImageJ).
The positions of the electrode needles are independently controlled using two sets of patch-clamp micromanipulators (Narisige, NMN-21).
Observations of prepared LNT films were also made using a digital microscope (Omron, VC7700), a polarised microscope (Olympus, BX51), and a field emission scanning electron microscope (JEOL, JSM-7300F) at 3 kV.

%%%%%%%%%%%%%%%%%%%%%%%%%%%%%5

\section*{Supplementary Discussion}
\subsection*{1. Length distribution of LNTs}

To measure the length variation of dispersed LNTs in the same suspension as that used for film fabrication, we applied a high-frequency AC electric field using the following method that has been developed in Ref. [33].
An AC field of 0.2 kV/cm at a frequency of 4 MHz was applied to a 30-$\mu$l aliquot via two vertical pairs of aluminium electrodes placed on a cover slip.
We stained the LNTs with the fluorescence dye 4',6-diamidino-2-phenylindole (DAPI) to visualise the individual components in sharp contrast.
The orientation of the LNTs can be switched orthogonally by a change in the active electrode pair, while dielectrophoretic assembly at 4 MHz hardly occurs, as inferred from Supplementary Figure S6(b).
It was observed that a considerable number of LNTs responsive to the AC electric field individually and simultaneously rotated from alignment to the vertical.

We verified the linear increase in rotation angle with time.
Hence, the angular velocity $\omega$ was evaluated from the slope of the rotation angle versus time, where $\omega=\pi/(2T)$, using a required time $T$ to rotate through an angle of $\pi/2$ radians.
Considering the balance equation between the exerted torque and the hydrodynamic force caused by water viscosity, it is found that $\omega$ noticeably decreases as $\omega\propto L^{-3}$, where $L$ denotes the length of the major axis of a LNT mimicked by a prolate spheroid.
Because $\omega$ is a sensitive indicator of $L$, we obtained the length distribution based on the previous procedure, which is displayed as a histogram in Supplementary Figure S1.
The histogram profile has one maximum and a tail in the longer-length region, and the mean length is approximately 6.1 $\mu$m.
Though there exists a small portion of LNTs with lengths longer than 10 $\mu$m (see Supplementary Figure S1), the 10-$\mu$m electrode gap and the off-chip position of the electrodes result in the exclusion of the long sedimentary LNTs and of random bundles from the constituents of the LNT array film.

\subsection*{2. Assembly processes of carboxylate-modified polystyrene (CM-PS) spheres}

Here, we describe extra data on the assembly processes of spherical colloids, CM-PS particles given in Supplementary Figures S2 to S4.
First, we look at the evolution of the total number $N(t)$ in Supplementary Figure S2(a), indicating that $N(t)$ is proportional to $t$, irrespective of the electric field strength.
The straight lines fitted to the data in Supplementary Figure S2(a) provide the slopes, which we have designated as the assembly number rate per second: $V_N=dN(t)/dt$.
The resulting $E$-dependence of $V_N$ has been shown in Figure 4(b).
From Supplementary Figure S6(a), on the other hand, we can see that $V_N$ is a decreasing function of $f$ in the range $100 \mathrm{kHz}\leq f\leq 700 \mathrm{kHz}$ while maintaining the strength $E=3.3$ kV/cm.

We also measured the evolution of the occupied area covering both the inside and outside of the rectangular region.
Supplementary Figure S2(b) shows that $P_A(t)$ increases and crosses 100 $\%$ at $t_f$, when the rectangular area is filled with colloidal particles prior to reaching $N_{\mathrm{max}}=7.4\times 10^2$.
From Supplementary Figure S2(a), it is found that the total numbers at $t_f$ are identical, independent of $E$: $N(t_f)=4.9\times 10^2$ for all of the different field strength magnitudes;
the constancy of  $N(t_f)$ means that the layer number and the accumulation degree are invariant with respect to $E$ at $t_f$ due to a uniform spread of collected particles in the plug-in system.
Because $V_N$ reads that $V_N=N(t_f)/t_f$, the constancy of $N(t_f)$ means that the $E$-dependence of $1/t_f$ accords with that of $V_N$.

Supplementary Figure S3 demonstrates a time series of optical micrographs, representing the assembly process of spherical colloids, CM-PS particles.
The optical micrograph on the leftmost side of Supplementary Figure S3 shows that colloids start to accumulate around the pointed electrode heads due to dielectrophoresis toward the needle tips, where the electric field gradient is obviously the largest.
The sequential images in Supplementary Figure S3 indicate that the colloidal clusters around both heads are pumped from both sides into the inside of the rectangular region as marked in Figure 1(b).

Lastly, Supplementary Figure S4 provides the height effect on the assembly courses in terms of $V_N$.
The height $h=0$ is determined by the contact position as follows:
lifting the parallel electrodes off the substrate at an arbitrary height, the needles were gradually lowered until they reached the bottom, where the electrodes were in contact with the cover slip.
Then, we can measure $h$ by vertically controlling the electrode position using the micromanipulators.
The maximum height $h_{\mathrm{max}}=100\,\mu\mathrm{m}$ is close to the air-water interface, determined by the contact angle of a water droplet on the substrate where Supplementary Figure S4 quantifies the $h$-dependence of $V_N$ for $E=3.3$ kV/cm and $f=100$ kHz.
Comparison between the optical micrographs on the right side of Supplementary Figure S4 also indicates that the assembly degree of colloidal arrays is reduced as the height $h$ from the substrate in an assembly is decreased.
At $h=0$, we observed that a solvent flow, such as electro-osmosis, in the vicinity of the glass substrate hinders the stable trapping of spherical colloids.

\subsection*{3. Assembly processes of LNTs}

We have adjusted the relative length of the electrode gap compared with the LNT length to create uniform films of parallel LNT arrays, or nanocapillary films. 
The gap length can reach a minimum width of 6-$\mu$m, equal to the mean LNT length where, theoretically, the channel configuration appears to be optimised;
however, we set the width to be 10 $\mu$m, longer than 6 $\mu$m, to avoid dielectrophoretic jamming around the electrode tips.
Consequently, the electric moulding has produced homogeneous LNT films, and the uniform assembly of LNTs validates the constancy of $N(t_f)$ that has been assumed in corroborating the result of Supplementary Figure S5(a) as follows.

Supplementary Figure S5(a) shows four plots of $P_A(t)$ for the field strengths $E=0.50,\,1.0,\,2.0,\,\mathrm{and}\,3.0\,\mathrm{kV/cm}$ at $f=300$ kHz.
From Supplementary Figure S5(a) as well as Supplementary Figure S2(b), we can see that the fraction $P_A(t)$ rises faster as $E$ increases;
so $t_f$ decreases with an increase in $E$.
Instead of counting individual LNTs, $1/t_f$ can clarify the $E$-dependence of $V_N$ assuming that $N(t_f)$ is independent of $E$, similar to the case of CM-PS particles.
Figure 4(d) displays the log-log plot of $1/t_f$ versus $E$, where the experimental results are found to satisfy the relation $1/t_f\sim E^2$, which is represented by the straight line.
So, the reference rate $1/t_f$ for LNTs holds the dielectrophoretic property as that for CM-PS particles does, thereby validating that in both CM-PS and LNT suspensions $V_N\propto E^2$.

Furthermore, we clarified the frequency dependence of the total assembly degree for LNTs using the averaged grey-scale over the rectangular area because there are some cases where $t_f$ is not directly related to $V_N$, such as the following:
at the higher frequencies of 500 kHz and 700 kHz, we observed that the accumulation of LNTs started after applying an external field for minutes (see Supplementary Fig. S5(b)), indicating that $V_N\propto t$ is violated.
Hence, the assembly courses of LNTs were assessed not by the occupied-area fraction $P_A(t)$, but by the normalised mean darkness $G(t)$ that measures transmitted light through the assembly films, where $G(t)$ has been obtained from the average of the grey-scale over the rectangular region, which was normalised to satisfy $G(t)\rightarrow 1$ for black objects.
In Figure S6(b), the total degree of gathering LNTs is expressed by $G(t_r)$ at a reference time $t_r=10\,\mathrm{min.}$ when $E=2$ kV/cm.
From Figure S6(b), it is found that $G(t_r)$ decreases sharply with an increase in $f$ in the range $100\,\mathrm{kHz}\leq f\leq 700\,\mathrm{kHZ}$, which is quite analogous to the $f$-dependence of $V_N$ for CM-PS particles in Supplementary Figure S6(a).

\subsection*{4. Evaluating the mean height of a colloidal assembly}.

Let $w_{\mathrm{sphere}}$, $\phi$, and $H$ be the spherical volume of a CM-PS particle, the volume fraction of an electrical assembly, and the mean height of an electrical box, respectively.
The volume fraction $\phi(t)$ as a function of the duration time $t$ is defined as
   \begin{equation}
   \phi(t)=\frac{N(t)w_{\mathrm{sphere}}}{A_{\mathrm{max}}H}.
   \label{fraction}
   \end{equation}
While we observed a hexagonal order in the first layer (see Fig. 4(a)), the electrode needles cannot maintain the 3D crystalline order, and it is plausible to assume that $\phi$ reaches a maximum $\phi_{\mathrm{RCP}}$ at a random close packing.
Hence, eq. (\ref{fraction}) implies the inequality, with respect to $H$,
   \begin{equation}
   H\geq
   \frac{N(t)w_{\mathrm{sphere}}}{A_{\mathrm{max}}\phi_{\mathrm{RCP}}}\equiv H_{\mathrm{min}}.
   \label{rcp-minimum}
   \end{equation}
When we use $\phi_{\mathrm{RCP}}=0.64$, the value of hard spheres in the literature (e.g., see Ref. [51]), and when we calculate the volume $w_{\mathrm{sphere}}$ using $w_{\mathrm{sphere}}=(4\pi/3)a^3$ and a spherical radius $a=1.5\,\mu\mathrm{m}$, the mean height $H_{\mathrm{min}}$ at the minimum is
   \begin{equation}
   H_{\mathrm{min}}=\frac{N(t_f)w_{\mathrm{sphere}}}{A_{\mathrm{max}}\phi_{\mathrm{RCP}}}=4.0\,\mu\mathrm{m}
   \label{first-layer}
   \end{equation}
at completing the first layer. The terminal height is given by
   \begin{equation}
   H_{\mathrm{min}}=\frac{N_{\mathrm{max}}w_{\mathrm{sphere}}}{A_{\mathrm{max}}\phi_{\mathrm{RCP}}}=6.1\,\mu\mathrm{m},
   \label{complete-height}
   \end{equation}
using the maximum total number $N_{\mathrm{max}}$.
While the former height (\ref{first-layer}) is comparable to the 3 $\mu$m diameter of the spheres present, the latter (\ref{complete-height}) is consistent with the thickness of LNT films.

\subsection*{5. Time dependencies of dielectrophoretic assembly courses}.

As explained in the literature (e.g. see Refs.  [50] and [52]), dielectrophoresis of colloidal particles is distinguished from electrophoresis by the following features.
First, dielectrophoresis is caused by the gradient of DC or AC electric field, in contrast with electrophoresis, which occurs even under a uniform field.
Moreover, while only charged objects undergo electrophoresis, any polarisable object, charged or neutral, experiences the dielectrophoretic force $F_{\mathrm{DEP}}$ given by
   \begin{eqnarray}
   F_{\mathrm{DEP}}&=&2\pi\epsilon_0\epsilon^*_m(0)a^3\mathrm{Re}[K^*(2\pi f)]\left|\nabla E\right|^2,\\
   \mathrm{Re}[K^*(2\pi f)]&=&\frac{\epsilon^*_p(f)-\epsilon^*_m(f)}{\epsilon^*_p(f)+2\epsilon^*_m(f)}.\label{cm}
   \end{eqnarray}
Here, $\epsilon_0$ is the vacuum permittivity, $\epsilon^*_m(f)$ is the complex dielectric constant of the solvent at an applied frequency $f$, $\epsilon^*_m(0)$ is the DC component of $\epsilon^*_m(f)$, $\epsilon^*_p(f)$ is the complex dielectric constant of a colloid at $f$, $a$ is the radius of a spherical colloid,  $E$ is the root mean square magnitude of the electric field, and $\mathrm{Re}[K^*(2\pi f)]$ corresponds to the real part of the Clausius-Mossotti (CM) factor as a function of $f$.

The resulting dielectrophoresis produces a hydrodynamic force exerted in the opposite direction of the dielectrophoretic velocity $v_{\mathrm{DEP}}$.
The balance equation between the dielectrophoretic and hydrodynamic forces gives
   \begin{equation}
   F_{\mathrm{DEP}}=6\pi\eta av_{\mathrm{DEP}},
   \label{SE}
   \end{equation}
where the right-hand side of the equation arises from the Stokes-Einstein relation with $\eta$ denoting the water viscosity.
Combining eqs. (\ref{cm}) and (\ref{SE}), we have
   \begin{equation}
   v_{\mathrm{DEP}}\propto \mathrm{Re}[K^*(2\pi f)]\left|\nabla E\right|^2.
   \label{v-dep}
   \end{equation}
Because the absolute value of the electric field gradient is supposed to be proportional to the electric field magnitude $E$, neglecting any non-linear effects, eq. (\ref{v-dep}) indicates that $v_{\mathrm{DEP}}$ is proportional to both $E^2$ and to the real part of the CM factor.

We relate the total assembly number $N(t)$ of colloids to the dielectrophoretic velocity $v_{\mathrm{DEP}}$ using the following conservation law:
   \begin{equation}
   \frac{\partial \rho({\bf r},t)}{\partial t}=-\nabla\cdot\rho({\bf r},t){\bf v}({\bf r},t),\label{conservation}
   \end{equation}
where $\rho({\bf r},t)$ is the positionally dependent colloid concentration and ${\bf v}({\bf r},t)$ represents a velocity vector at position ${\bf r}$.
Integrating both sides of eq. (\ref{conservation}) over the spatial region, we obtain the relation
   \begin{equation}
   \frac{dN(t)}{dt}=-\int d{\bf r}\nabla\cdot\rho({\bf r},t){\bf v}({\bf r},t),
   \label{int-c}
   \end{equation}
with respect to the total number $N(t)=\int d{\bf r}\rho({\bf r},t)$ of colloids entering the rectangular area (see Fig. 1(b)).
Assigning the rectangular frame to the boundary surface of the integral region, $N(t)$ in eq. (\ref{int-c}) is the total particle number.

Correspondingly, we transform the right-hand side to a surface integral with respect to the aforementioned boundary:
   \begin{equation}
   -\int d{\bf r}\nabla\cdot\rho({\bf r},t){\bf v}({\bf r},t)
   =-\rho_0\int dS\,{\bf v}({\bf r},t)\cdot{\bf n},
   \label{green}
   \end{equation}
where ${\bf n}$ denotes the unit vector vertical to the boundary, as usual.
In expression (\ref{green}), $\rho_0$ corresponds to the boundary density, equal to the bulk concentration in the first approximation, and the vertical component ${\bf v}\cdot{\bf n}$ of the boundary velocity is mainly determined by dielectrophoresis across the regions near the electrode tips.

Combining eqs. (\ref{v-dep}), (\ref{int-c}), and (\ref{green}), we find the following:
 \begin{itemize}
\item[] (i) The assembly rate $V_{N}=dN(t)/dt$ satisfies the proportionality $V_{N}\propto v_{\mathrm{DEP}}\propto E^2$ as long as $\rho_0$ is independent of $E$.
\item[] (ii) $V_N$ depends on $f$ through the $f$-dependence of $\mathrm{Re}[K^*(2\pi f)]$, which has been found to be a decreasing function of $f$ for both LNTs and CM-PS particles (see Supplementary Fig. S6) as is the case with the spectra of $\mathrm{Re}[K^*(2\pi f)]$ for various colloids [52].
\end{itemize}
\end{document}